# Efficient terahertz electro-absorption modulation employing graphene plasmonic structures

Berardi Sensale-Rodriguez[a], Rusen Yan, Mingda Zhu, Debdeep Jena, Lei Liu, and Huili Grace Xing[b].

*Department of Electrical Engineering, University of Notre Dame, Indiana 46556 USA.*

**Abstract**

We propose and discuss terahertz electro-absorption modulators based on graphene plasmonic structures. The active device consists of a self-gated pair of graphene layers, which are patterned to structures supporting THz plasmonic resonances. These structures allow for efficient control of the effective THz optical conductivity, thus absorption, even at frequencies much higher than the Drude roll-off in graphene where most previously proposed graphene-based devices become inefficient. Our analysis shows that reflectance-based device configurations, engineered so that the electric field is enhanced in the active graphene pair, could achieve very high modulation-depth, even ~100%, at any frequency up to tens of THz.

[a] bsensale@nd.edu

[b] hxing@nd.edu

The terahertz (THz) frequency range has recently become a very dynamic area of scientific research due to its important applications in diverse disciplines such as astronomy, biological and chemical sensing, communications, security, etc.[1]. However, there is still a lack of electrically-driven semiconductor devices capable of efficiently manipulating THz waves, such as active filters, modulators, switches, etc. Although several proof-of-concept prototypes were proposed and demonstrated over the past decade (e.g. Ref.[2-5]), these devices so far have presented severe tradeoffs between insertion loss and modulation depth[6].

Owing to its two-dimensional nature, which leads to unprecedented integration possibilities and remarkable mechanical, electrical, and optical properties, graphene has become a promising material for future electronics. Although in the infrared/visible range its optical absorption is only a few percent and scarcely controllable, optical conductivity dramatically increases in the terahertz range leading to the possibility of electrical control of terahertz absorption[7]. We have recently demonstrated that single layer graphene is capable of efficiently tuning THz transmission meanwhile introducing negligible insertion loss[6,8-9]. However these devices can only operate efficiently up to the Drude roll-off frequency of the optical conductivity. In the upper THz range (i.e. frequencies above 2 THz) although the magnitude of the optical conductivity of graphene quickly decreases because of its Drude dispersion, intense plasmonic effects can occur. Room-temperature plasmonics were demonstrated in periodic graphene structures leading to the possibility of strongly absorbing THz radiation. Passive filters and polarizers were recently demonstrated employing this phenomenon[10]; these structures are also able of modulating THz transmittance when gated[11]. In general, by combining active graphene layers with other passive structures augmenting the intensity of the electric field in graphene (e.g. reflectors), the control over THz waves can be greatly enhanced.

In this paper, we propose plasmonic electro-absorption modulators of THz reflectance, capable of exhibiting superior performance than the prior art. By patterning the active graphene layers into micro-ribbons thus taking advantage of the plasmonic effects, as well as electric field enhancement in the active regions of the device by a back reflector, we show that it is possible to achieve very high modulation depth at any frequency in the THz range.

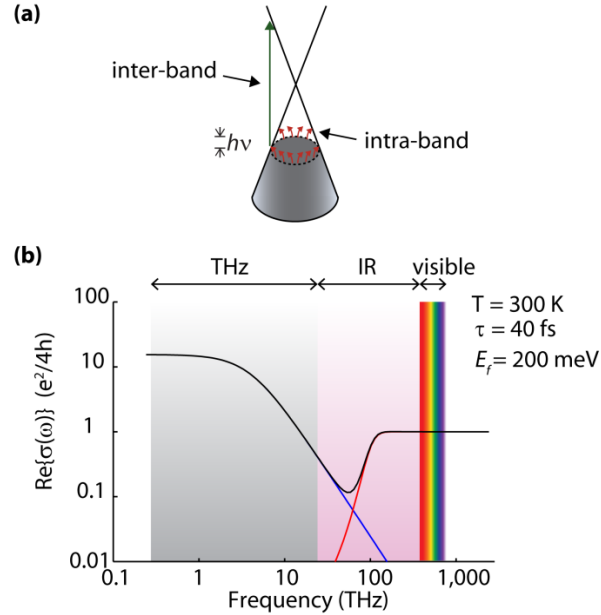

FIG. 1. (a) Optical transitions in graphene. (b) Real part of the optical conductivity in graphene versus frequency.

Optical conductivity in graphene can be calculated from the contributions of inter and intra-band transitions as shown in Fig. 1a. However, as pictured in Fig. 1b, in the upper THz/far-IR range, the optical conductivity is small since neither intra-band nor inter-band transitions are strong. Owed to this observation, electrical tuning of optical conductivity becomes challenging in this frequency range.

Patterned graphene or other electronic structures can be platforms for collective oscillations of charge carriers, the so-called plasmons. The effective conductivity in plasmonic structures is related to its DC conductivity, carrier scattering time as well as other geometric parameters. For instance, the effective sheet conductivity of planar arrays of conducting disks[12] can be described by the following closed form:

$$\sigma(\omega) = f\sigma_{DC}\frac{i\omega\tau}{(\omega^2 - \omega_p^2) + i\omega\tau}, \qquad (1)$$

where $\sigma_{DC}$ is the DC electrical conductivity of graphene, $f$ is the filling factor, $\omega$ is angular frequency, $\omega_p$ is the plasmon resonance frequency, and $\tau$ is the carrier momentum scattering time. The plasmon resonance frequency is proportional to $\sigma_{DC}^{1/2}$ (Ref.[10-12]). However, since $\sigma_{DC}$ of graphene is proportional to $n_s^{1/2}$, resulting from its linear $E$-$k$ dispersion, $\omega_p$ is thus proportional to $n_s^{1/4}$, which contrasts with the $n_s^{1/2}$ dependence in conventional semiconductors. As seen in Eq. (1), at the plasmon resonance frequency, the optical conductivity thus absorbance can have the same strength as below the Drude roll-off, i.e. $\sigma(\omega_p) = f\sigma_{DC}$. Since $\sigma_{DC}$ in graphene can be easily tuned by field effect, plasmonic structures allow to extend the tunable range of THz components to frequencies higher than that limited by the Drude roll-off frequency (carrier momentum scattering time). Based on these properties, Yan *et al* demonstrated passive notch filters and polarizers employing arrays of graphene disks and ribbons, respectively[10]. Moreover, voltage tunable devices were also reported employing an array of graphene ribbons on $SiO_2$/Si gated by ionic gel[11]. However, the attained maximum absorption at the plasmonic peak was only ~14%, which is consistent with the maximum achievable DC conductivity in typical CVD graphene. In analogy to what we have recently demonstrated in unpatterned graphene THz modulators for frequencies below the Drude roll-off, plasmonic effects can also be combined with other passive and field concentrating structures to enhance modulation[6].

A simple way to enhance electric field in the active graphene layers is to employ reflectance based structures[6]. Let's consider the device structure depicted in Fig. 2, which consists of a self-gated graphene pair[13] patterned in ribbons lying on top of a substrate with a metal reflector deposited on the back side.

For the self-gated graphene pair (active region of the device) we will assume that under zero-bias, the Fermi level of both graphene layers is at Dirac point (Fig. 2b). Thus the overall effective electric conductivity of the structure under zero bias is given by: $\sigma_{DC,total}(0V) = 2\sigma_{DC,min}$, where $\sigma_{DC,min}$ is the minimum DC conductivity of graphene ($\sigma_{DC,min} \sim 0.15$ mS as discussed in Ref.[14]). If a finite bias is applied, carriers of opposite type accumulate in each of the graphene layers, leading thus to an increase in the overall effective electric conductivity of the structure.

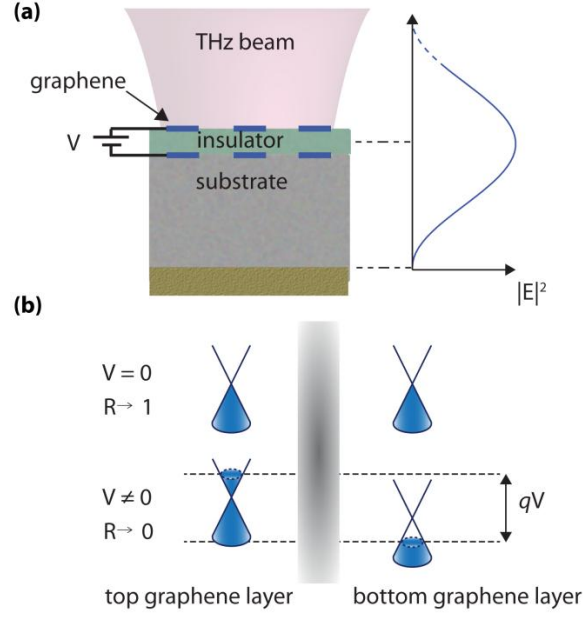

FIG. 2. (a) Schematic of the device, and electric field distribution when the substrate thickness and plasmonic resonance are matched to an odd multiple of a quarter wavelength of the THz wave. (b) Operation mechanism of the self-gated graphene pair.

Employing the transfer matrix formalism[8], the reflectance in this structure can be written as:

$$R = \left| \frac{(n_s + 1 - Z_0\sigma)e^{-j\varphi} + (n_s - 1 + Z_0\sigma)e^{j\varphi}}{(n_s - 1 - Z_0\sigma)e^{-j\varphi} + (n_s + 1 + Z_0\sigma)e^{j\varphi}} \right|^2, \qquad (2)$$

where $\varphi = 2\pi t / \lambda$, $t$ is the substrate thickness, $\lambda$ the THz wavelength in the substrate, $n_s$ the substrate refractive index (i.e. $n_s = 3.42$ for Si), $Z_0 = 377$ Ω is the vacuum impedance, and $\sigma$ is the graphene conductivity.

The field enhancement in these structures can be understood as follows. If the substrate optical thickness is chosen as an odd multiple of the quarter wavelength at the plasmon resonance in graphene ($t = (2N-1) \lambda_p / 4n_s$ where $N = 1,2,...$ is an integer, and $\lambda_p$ is the plasmon resonance wavelength in vacuum), then the electric field ($E$) exhibits a maximum at the active graphene pair and a node at the back metal (i.e. $E$ is null). The maximum field strength is 4X of the field strength of the propagating plane wave, thus leading to enhanced absorbance (Fig. 2a). Reflectance at this frequency can be written as:

$$R(\omega_p) = \left|\frac{1 - Z_0 \sigma(\omega_p)}{1 + Z_0 \sigma(\omega_p)}\right|^2, \tag{3}$$

Substituting Eqn. (1) into Eqn. (3), we obtain:

$$R(\omega_p) = \left|\frac{1 - Z_0 f \sigma_{DC,total}}{1 + Z_0 f \sigma_{DC,total}}\right|^2, \tag{4}$$

Considering a 50% filling factor (i.e. $f = 0.5$), and assuming the hole and electron mobility in graphene to be the same (so $\sigma_{DC,total} = 2\sigma_{DC}$ where $\sigma_{DC} = \sigma_{DC\ top\ layer} = \sigma_{DC\ bottom\ layer}$), the modulation depth achievable at the plasmon resonance, $MD = |R(0V) - R(V)| / R(0V)$, is thus:

$$MD(\%) \approx 100\left(1 - \left|\frac{1 - Z_0 \sigma_{DC}(V)}{1 + Z_0 \sigma_{DC}(V)}\right|^2\right), \tag{5}$$

Employing Eqn. (2), we calculated modulation depth for the graphene geometry and material parameters reported in Ref.[11] (i.e. ribbon width = 2 μm, $f = 0.5$). Shown in Fig. 3 is the calculated modulation depth versus frequency for different substrate thicknesses using a conductivity swing of graphene extracted from the experimental data in Ref.[11]. When the substrate thickness is an odd multiple of a quarter wavelength of the plasmonic resonance, in accordance with the previously described theory, the modulation depth in this device approaches 70%. The improvement with respect to the transmittance modulator ($MD \sim 14\%$) is owed to the aforementioned field enhancement.

The insertion loss ($IL$), assuming a minimum achievable graphene conductivity of 0.15mS, is 18% (or 0.8dB). This value is much lower than what is attainable in other modulator technologies achieving similar modulation depths[2-5]. Such a low $IL$ is also consistent with our previous experimental studies in large area graphene THz reflectance modulators[6]. In terms of the THz wave being modulated, its frequency can be designed with the proper choice of the graphene ribbon width ($W$) since $\omega_p$ is proportional to $1 / W^{1/2}$ (Ref.[11]), e.g. 10 THz with a ribbon width of 200 nm.

Finally, a very interesting observation can be derived from Eqn. (5): ~100% modulation depth can be achieved if graphene conductivity can be tuned to be $\sigma_{DC} = 1/377\ \Omega = $ ~2.7mS.

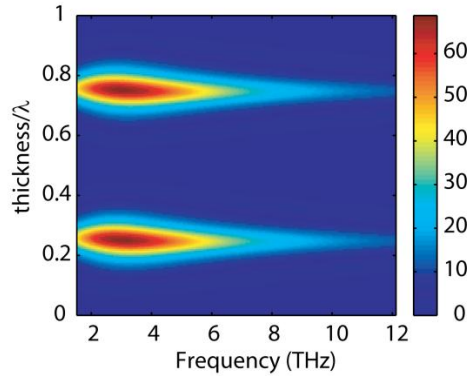

FIG. 3. Contours of modulation depth versus frequency and ratio between substrate thickness and THz wavelength; high modulation is possible due to field enhancement and plasmonic absorption.

In synthesis, we have presented THz wave reflectance modulators exhibiting superior performance due to electric field enhancement in the active regions by a back reflector. By patterning the active graphene layers into micro-ribbons, taking advantage of plasmonic effects, we show that is possible to achieve very high modulation depth, even ~100%, at any frequency up to tens of THz.

The authors acknowledge the support from NSF (CAREER ECCS-084910, ECCS-1202452), the Center for Advanced Diagnostics and Therapeutics (AD&T) and the Center for Nanoscience and Technology (NDnano) at the University of Notre Dame.

**References**

[1] M. Tonouchi, Nature Photon. **1**, 97 (2007).

[2] T. Kleine-Ostmann, P. Dawson, K. Pierz, G. Hein, and M. Koch, Appl. Phys. Lett. **84**, 3555 (2004).

[3] P. Kuzel, F. Kadlec, J. Petzelt, J. Schubert, and G. Panaitov, Appl. Phys. Lett. **91**, 232911 (2007).

[4] H. T. Chen, W. J. Padilla, M. J. Cich, A. K. Azad, R. D. Averitt, and A. J. Taylor, Nature Photon. **3**, 141 (2009).

[5] H. T. Chen, H. Lu, A. Azad, R. Averitt, A. Gossard, S. Trugman, J. O'Hara, and A. Taylor, Opt. Express **16**, 7641 (2008).

[6] B. Sensale-Rodriguez, R. Yan, S. Rafique, M. Zhu, W. Li, X. Liang, D. Gundlach, V. Protasenko, M. M. Kelly, D. Jena, L. Liu, and H. G. Xing, Nano Lett. **12**, 4518 (2012).

[7] J. M. Dawlaty, S. Shivaraman, J. Strait, P. George, M. Chandrashekhar, F. Rana, M. G. Spencer, D. Veksler, and Y. Chen, App. Phys. Lett. **93**, 131905( 2008).

[8] B. Sensale-Rodriguez, T. Fang, R. Yan, M. M. Kelly, D. Jena, L. Liu, and H. G. Xing, Appl. Phys. Lett. **99**, 113104 (2011).

[9] B. Sensale-Rodriguez, R. Yan, M. M. Kelly, T. Fang, K. Tahy, W. S. Hwang, D. Jena, L. Liu, and H. G. Xing, Nat. Commun. **3**, 780 (2012).

[10] H. Yan, X. Li, B. Chandra, G. Tulevski, Y. Wu, M. Freitag, W. Zhu, P. Avouris, and F. Xia, Nat. Nano. **7**, 330 (2012).

[11] L. Ju, B. Geng, J. Horng, C. Girit, M. Martin, Z. Hao, H. A. Bechtel, X. Liang, A. Zettl, Y. R. Shen, and F. Wang, Nat. Nano. **6**, 630 (2011).

[12] S. J. Allen, H. L. Stormer, and J. C. M. Hwang, Phys. Rev. B **28**, 4875–4877 (1983).

[13] S. Koester, H. Li, and M. Li, Opt. Express **20**, 20330 (2012).

[14] N. M. R. Peres, and T. Stauber, Int. J. of Mod. Phys. B, **22**, 2529 (2008).